\documentclass[12pt]{article}

\usepackage{geometry}
\usepackage[round]{natbib}
\usepackage{graphicx}
\usepackage{authblk}
\usepackage{setspace}
\usepackage{amsmath,multirow,pbox}

\title{\vspace{-4ex}Inferring habitat quality and habitat selection using static site occupancy models.}

\author[a]{Philip T. Patton\thanks{ptpatton@ncsu.edu}}
\author[a]{Krishna Pacifici}
\author[a,b]{Jaime Collazo}
\affil[a]{\small{\emph{Program in Fisheries, Wildlife, and Conservation Biology, Department of Applied Ecology, North Carolina State University, USA}}}
\affil[b]{\small{\emph{U.S. Geological Survey, North Carolina Cooperative Fish and Wildlife Research Unit, North Carolina State University, USA}}}

\date{\vspace{-6ex}}

\begin{document}
	
\maketitle

\section*{Abstract}

When evaluating the ecological value of land use within a landscape, investigators typically rely on measures of habitat selection and habitat quality. Traditional measures of habitat selection and habitat quality require data from resource intensive study designs (e.g., telemetry, mark--recapture, or multi--season point counts). Often, managers must evaluate ecological value despite only having data from less resource intensive study designs. In this paper, we use occupancy data to measure habitat quality and habitat selection response for the Puerto Rican Vireo, an endemic songbird whose population growth is depressed by brood parasitism from the Shiny Cowbird. We were interested in how vireo habitat quality and vireo habitat selection response varied among three land uses---forest, shaded coffee plantations, and sun coffee---in a coffee growing region in Puerto Rico. We estimated vireo occupancy probability as a measure of habitat selection, and the probability of cowbird occurrence given vireo presence as a measure of habitat quality. To estimate the latter, we explored different ways of incorporating host information into joint occurrence models. These included the conditional occurrence model \citep{waddleEA2010}, a two-stage hierarchical Bayesian model that propagates the joint uncertainty from the vireo and the cowbird through to estimates of co-occurrence, and a commonly used model that includes the na\"ive occurrence of the host as a covariate. We assessed the predictive performance of each model using posterior predictive checks, a pseudo--$R^2$, and DIC. Vireos preferentially selected forested sites and shaded coffee sites over sun coffee sites. By our measure of habitat quality, either type of coffee plantation was of poor quality, and the forested sites were high quality. This suggests that shade coffee may be an ecological trap for the vireo in our study area. The two--stage model performed best by our measures of predictive ability, thus showing that the cowbird may not occur independently of the vireo. Vireo population dynamics in our study area may benefit from having large amounts of forest relative to coffee plantations. Incorporating species interactions into occupancy models has the potential to improve monitoring for conservation.
   
\section*{Keywords}
occupancy; detection; brood parasitism; Bayesian model selection; \emph{Vireo latimeri}; \emph{Molothrus bonariensis}; shade coffee

\section*{Introduction}

Patterns in occurrence can effectively inform management practice \citep{zipkinEA2009,conroyEA2011} when the probability of occurrence is positively correlated with habitat quality. Occupancy probablity and habitat quality will not be correlated when organisms make suboptimal decisions in selecting habitats to occupy, such that they avoid high--quality habitats \citep[perceptual errors;][]{gilroysutton2007} or preferentially choose poor quality habitats \citep[ecological traps;][]{battin2004,robertsonEA2013}. These phenomena occur when species are slow to respond to a rapidly changing environment, such as increasing densities of nest predators, parasites, or diseases \citep{robertsonEA2013}. For example, grassland birds still nest in patches of grassland that have long been unviable because of high densities of Brown--headed Cowbirds \citep{johnsontemple1990}. Both ecological traps and perceptual errors can drive species to extinction. Therefore, understanding how habitat selection response and habitat quality vary by site is critical for monitoring for conservation. Managers who ignore habitat selection and habitat quality could unwittingly propagate low quality habitats that organisms preferentially select (i.e., ecological traps), leading to meta--population extinction. Unfortunately, standard measures of habitat quality (e.g., survival) and habitat selection (e.g., resource use) require expensive field studies that decision makers often cannot afford. In this paper, we use detection/non--detection data, which are comparatively cheap to collect, and static site occupancy models \citep{mackenzieEA2002} to understand how habitat quality and habitat selection vary by site in a tropical agricultural landscape.

Shade coffee---shorthand for the practice of farming coffee under 15 to 35\% canopy cover---is thought to be beneficial to bird species, but may actually be an ecological trap. The bird friendly reputation is based on many studies that found elevated individual species abundances and community species richness relative to sun coffee plantations \citep{perfectoEA1996,philpottEA2008,borkhatariaEA2012,jhaEA2014}. However, certain authors speculate that shade coffee plantations might actually be an ecological trap for certain songbirds \citep{buechleyEA2015}. The vegetative characteristics of shade coffee plantations are similar to those of edge habitats, which are known to be an ecological trap for woodland songbirds and some forest interior birds \citep{battin2004,weldonhaddad2005,kingEA2009,newmarkstanley2011}. These birds establish territories and nests within the vegetation at the forest edge. These nests are at greater risk of predation or parasitism from predators or parasites that inhabit the open habitats next to to the forest edge. Both habitats---shade coffee and forest edge---are semi--open and often border open land uses (e.g., residential areas, sun--coffee plantations, or other agriculture). Open land uses may supply nest predators or nest parasites to shade--coffee plantations, thereby increasing nest failure rates in the plantations, and depressing population growth. Going beyond individual species occupancy or community richness to understand habitat quality would be beneficial to managers considering proliferating shade--coffee. We are unaware, however, of any studies providing evidence for whether shaded plantations are an ecological trap for songbirds.

To explore habitat selection and habitat quality in shaded coffee, we consider an example of an endemic songbird and an invasive brood parasite in a coffee growing region in Puerto Rico. The Shiny Cowbird (\emph{Molothrus bonariensis}) was introduced to Puerto Rico sometime in the mid--$20^{\mathrm{th}}$ century \citep{postwiley1977a}. After introduction, the cowbird proliferated throughout the island, parasitizing a multitude of host species \citep{wiley1988}. This parasitism depresses the reproductive rates of Puerto Rican Vireos (\emph{Vireo latimeri}) in Gu\'anica State Forest, where the two species co--occur with high probability \citep{woodworth1997,irizarryEA2016}. Matrix models that include these low reproductive rates project vireo population growth rates below one \citep{woodworth1999}. These projections could explain the observed declines in counts of vireos in the forest \citep{faaborgEA1997}. The overall status of the vireo, which only occurs in Puerto Rico, is uncertain \citep{birdlifeVireo}. While nest parasitism rates are high in the lowlands in and around Gu\'anica \citep{woodworth1999}, these rates are near zero in montane forests \citep{tossas2008}. Less is known about montane agricultural regions (i.e., the principle coffee growing region). Heightened understanding of habitat quality and habitat selection response in this coffee growing region could have important implications for vireo conservation and management. 

In this paper, we used detection/non--detection data to provide clues about habitat quality and habitat selection for the Puerto Rican Vireo in a coffee growing region in Puerto Rico. We used probability of occurrence of the vireo ($\text{Pr}(z^V),$ where $z$ is the unobserved occupancy state and $V$ denotes vireo) as a metric for landscape--level habitat selection. In theory, the probability of site occupancy corresponds to the probability that a site overlaps an individual's home range \citep{efforddawson2012}. Thus, differences in occupancy probability among land uses suggest differences in the distribution of home ranges, and how the species is selecting habitats at the landscape level (see \emph{Discussion}). Occupancy models provide accurate predictions of occurrence because they incorporate uncertainty created by the observation process \citep{mackenzieEA2002, mackenzieEA2006}. We used the probability of occurrence of cowbirds given vireo presence ($\text{Pr}(z^C | z^V),$ where $C$ denotes cowbird) as a measure of habitat quality. This measure is a surrogate for vireo reproductive rates because we know that the vireo is a preferred host of the cowbird \citep{wiley1988} and that vireo reproductive rates are depressed by cowbird parasitism \citep{woodworth1997,woodworth1999}.

The choice of model to estimate the conditional occurrence of cowbirds depends on whether the cowbird occurs independently of vireos. We hypothesized that cowbird occurrence may depend on vireo occurrence because, during the breeding season, the parasite may be more likely to occur at sites where their preferred host occurs. If so, then using occupancy models that incorporate interspecific interactions will perform better than single--species models \citep{mackenzieEA2004,waddleEA2010}. With estimates of the conditional occurrence of the cowbird and the occurrence of the vireo, we estimated the probability of co--occurrence by habitat ($\text{Pr}(z^C , z^V) = \text{Pr}(z^C | z^V)\text{Pr}(z^V)$) to synthesize habitat selection and habitat quality. 

\section*{Materials and Methods}

In this section, we detail how we generated posterior distributions for habitat specific occupancy probability for the vireo, and the probability of cowbird occurrence given vireo presence.

\subsection*{Study Area and Data}

\begin{figure}[h!]
	\caption{Map of the study area, including sample points and overlapping protected areas.}
	\label{fig:sa}
	\begin{center}
		\includegraphics{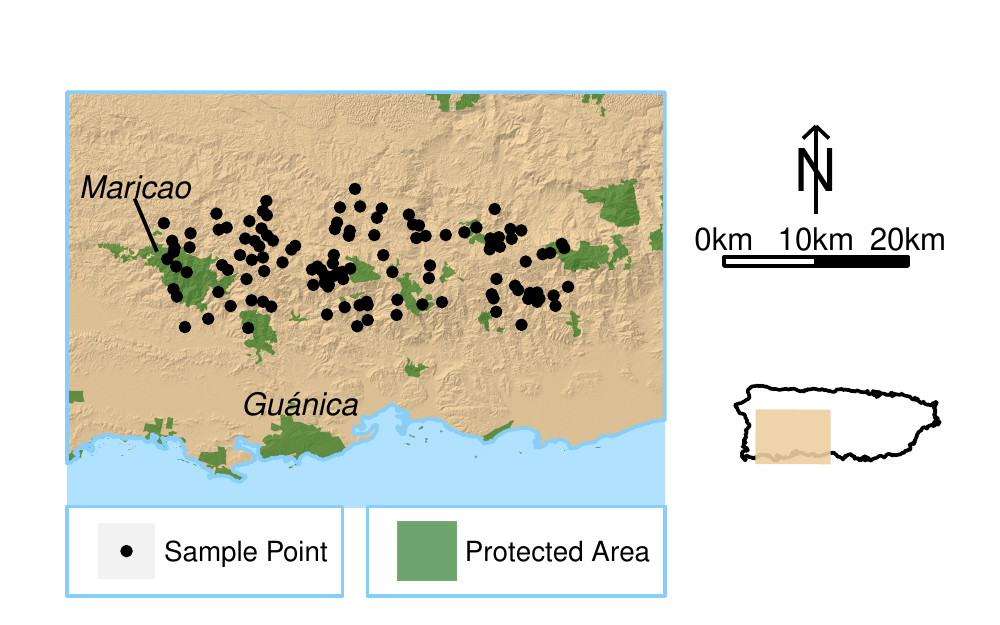}
	\end{center}
\end{figure}

We sampled a \(190 \mathrm{km}^2\) section of the western/central mountainous region of Puerto Rico (Figure \ref{fig:sa}). The study region ranges from mid-- to high--elevation, including the highest point in Puerto Rico, Cerro de Punta (1338 m). The area comprises multiple land uses, including forest, agriculture, and small urban centers (e.g. the municipalities of Adjuntas and Maricao). The agriculture is predominately coffee, citrus, and banana plantations. Much of the forest is considered upland tropical moist forest, or lowland tropical moist forest \citep{collazogroom2001}. Three large important protected landscapes---Maricao, Sus\'ua, and Monte Guilarte State Forests---overlap the study area. 

We surveyed 120 randomly selected sites, stratified by land use type (41, 37, and 42, sites in forest, sun coffee plantations, and shade coffee plantations). To assure independence among the sample points, they were spaced 500m apart. At each site, we conducted community level bird surveys from March to June in 2015. This time period is the peak of the breeding season for nearly the entire bird community \citep{collazogroom2001}. Almost every site was surveyed three times (although some were visited four times and others were visited two times) by one of two teams of observers---denoted here as observers A and B. The observers recorded the detection of any species seen or heard within a 50m radius over a 12--minute period. We used a 50m radius to minimize any heterogeneity in detection probability in the sample radius. Surveys took place from 30 minutes before sunrise to 10 AM. 

\subsection*{Habitat Selection}

We used habitat specific vireo occupancy probability as a surrogate for vireo habitat selection.  This surrogate was estimated with a single--season single--species occupancy model \citep{mackenzieEA2002}. We modeled heterogeneity in occurrence and the detection processes with a logit--linear model. Land use was the only covariate that we considered that could affect the occurrence process. Covariates representing indirect gradients \citep{austin2002}, such as elevation, longitude, and latitude, did not interest us for this analysis. Occurrence should vary among the three levels of land use---forest, shade coffee, and sun coffee---because vireos tend to be associated with forested habitats \citep{irizarryEA2016}. Observer was the only covariate considered to affect the detection process. Preliminary analysis showed that other covariates (e.g., date and time) did not enhance or hinder detection. The ability to detect individual species in community level bird surveys typically varies among observers. As stated above, one of two separate teams of observers conducted each survey. Each team, A and B, conducted 191 and 172 surveys. All told, we fit four occupancy models for vireos, each representing a possible combination of covariates: $\psi$(.)$p$(.), $\psi$(land use)$p$(.), $\psi$(.)$p$(observer), and $\psi$(land use)$p$(observer). 

We used DIC to select the model that best fit the data \citep{spiegelhalterEA2002}, and used this model to estimate habitat selection. Models with lower DIC's are considered to have greater predictive ability. Differences in DIC greater than 10 are thought to provide strong evidence that one model performs better than another model \citep[pp. 526]{gelmanhill2007}. In addition, we evaluated the degree of overlap among credible intervals for estimates of occurrence in each habitat. The best model was chosen based on its DIC and the significance of covariate effects.

\subsection*{Habitat Quality}

We used the probability of cowbird occurrence given vireo presence as surrogate for habitat quality. If the cowbird occurred independently of the vireo, this conditional occurrence probability could have simply equaled cowbird occurrence (i.e., $\text{Pr}(z^C | z^V) = \text{Pr}(z^C$)). However, we hypothesized that cowbirds would be more likely to occur where vireos occur. To test this hypothesis, and estimate our surrogate for habitat quality, we conducted a two step model building and selection process. The first step was identical to the vireo occurrence model building and selection process detailed in \emph{Habitat Selection}. By the end of this step, we had selected the best performing model among cowbird occurrence models---these four models are called the ``independent'' models in Table \ref{tab:mods}---that assumed that the cowbird occurred independently of vireos. We started the next step by taking this model and adding to it three different parameterizations for the relationship between the cowbird and vireo occurrence processes---these three models are called the ``dependent'' models in Table \ref{tab:mods}. After fitting these three models, which corresponded to three hypotheses about the relationship between the parasite and its host, we compared them to each other and the model that assumed no relationship using three model selection criteria: DIC, a pseudo--$R^2$ \citep{tjur2009}, and posterior predictive checking. Proceeding in this way promoted the efficiency of model fitting and selection without missing models that fit well. We detail this process below, beginning with a description of each ``dependent'' model.  

The first of these models is arguably the most common way to model species interactions with occurrence data \citep{wiszEA2013}. In this formulation, the observed (i.e., na\"ive) occurrence state of the vireo is treated as a site--level covariate affecting the occurrence of cowbirds, 

\begin{equation}
\label{eq:nai}
\mathrm{logit}(\psi^C_{j}) = \alpha^C_{\mathrm{land}(j)} + \alpha1*\mathrm{vireo}_j,
\end{equation} 
where $\text{vireo}_j$ is a site-level covariate denoting the detection of a vireo in at least one survey, and the superscript $C$ denotes cowbird. Throughout the rest of the paper, we refer to this model as the na\"ive covariate model or equation \ref{eq:nai} because the covariate is the na\"ive occurrence state of vireos. This model assumes that the occurrence of a vireo is known with certainty. 

To relax this assumption, we fit a two-stage hierarchical Bayesian model that addresses the joint uncertainty from the vireo and the cowbird occurrence processses,

\begin{equation}
\label{eq:z}
\mathrm{logit}(\psi^C_{j}) = \alpha^C_{\mathrm{land}(j)} + \alpha1*\tilde{z}^V_j,
\end{equation}
where the superscript $V$ denotes vireo. We call this the two--stage model because it requires initially developing a predictive model of vireo occurrence. In the second stage we use this model to generate a posterior predictive distribution of the occurrence of a vireo at each site. This posterior predictive distribution is used as a site--level covariate, $\tilde{z}^V_j,$ affecting the cowbird occurrence process. This model assumes that the effect of the presence of a vireo on cowbird occurrence is constant across land uses. 

The effect of vireos on cowbird occurrence could depend on the land--use. For example, cowbirds might preferentially occur in sun coffee plantations regardless of the presence of a host because sun coffee plantations provide food resources for this open--habitat dwelling insectivore. In contrast, cowbirds may only occur in forest sites where the vireo occurs because food resources for cowbirds are limited in forest. Thus, they probably only occur there to breed. To account for the possibility that the effect of vireo occurrence depends on land use, we used the conditional occurrence model developed by \cite{waddleEA2010}. This model is also a two--stage model, in that the vireo latent occurrence state is modeled independently of the cowbird occurrence state. In contrast, the cowbird occupancy state is conditioned on the occurrence state of vireos. Thus, the probability of occurrence of a cowbird is,

\begin{equation}
\psi^C = \text{Pr}(z^C=1) = 
\begin{cases}
\psi^{C|V} & \text{if } z^V = 1,\\
\psi^{C|\bar{V}} & \text{if } z^V = 0
\end{cases}
\end{equation}
where $\psi^{C|V}$ is the probability of occurrence for cowbirds given that a vireo is present and $\psi^{C|\bar{V}}$ is the probability of occurrence for cowbirds given that vireos are absent. Unlike \cite{waddleEA2010}, we modeled the detection of cowbirds and vireos separately, rather then explicitly conditioning the cowbird detection process on the vireo occurrence process. Preliminary analysis showed that the conditional detection model was uncompetitive by any measure. The full conditional occurrence model is

\begin{equation}
\label{eq:wad}
\text{logit}(\psi^C_j) = \alpha^{C|V}_{\mathrm{land}(j)}*\tilde{z}^V_j + \alpha^{C|\bar{V}}_{\mathrm{land}(j)}*(1 - \tilde{z}^V_j).
\end{equation} 

\subsubsection*{Model Selection}

To select the model that would be used to estimate our surrogate for habitat quality, we compared the fit and predictive ability of the model from step one of the process (i.e., the best performing ``independent'' model) and the three dependent models. Statisticians have developed a number of methods for evaluating the performance of Bayesian probability models \citep{hootenhobbs2015}. Practicioners typically choose the optimal method based on the desired use of the model (e.g., predicition) and the quantity of available data. We wanted to select a model based on its predictive ability despite having a relatively small data set. We chose three measures of fit and predictive ability: DIC, posterior predictive checks, and a pseudo--$R^2$  \citep{tjur2009}. This pseudo $R^2$, called the coefficient of discrimination, has many of the important properties of an $R^2$, and as such is an attractive way to estimate the explanatory power of a logistic regression model. The $R^2$ is calculated by predicting the mean probability of occurrence for sites with an estimated occurrence. Then, the mean probability of occurrence is computed for sites with an estimated absence. The difference of the two probabilities is the coefficient of discrimination. If the model is perfect, it will predict the mean probability of occurrence at sites with an occurrence to be 1.0, and 0.0 at sites without an occurrence, thus yielding an $R^2$ of 1.0. 

We also used DIC to arbitrate amongst these models. Using DIC to compare models that do and do not include the vireo detection/non--detection data (e.g., comparing the models in equations \ref{eq:nai} and \ref{eq:z}) would be improper because the inclusion of the vireo data increases the model's deviance. To make the estimates of DIC comparable across all models, we used the marginal likelihoods of both independent cowbird and vireo models to compute DIC for equations \ref{eq:nai} and \ref{eq:z} \citep{gelmanEA2013}.

Last, we used posterior predictive checks to evaluate the fit of each model \citep{gelmanEA2013,chambertEA2014}. The idea of the posterior predictive check is to select a quantity of particular ecological interest, then quantify how well a model predicts that quantity by comparing the predictions to observed values \citep{chambertEA2014}. In our case, the quantities of interest were the number of sites where the two species co-occur in forests, shade coffee, and sun coffee; and the number of sites where the vireo occurs by itself in forests, shade coffee, and sun coffee (six quantities in all, Figures \ref{fig:ppc} and \ref{fig:ppcHist}). The number of co--occurring sites in each habitat interested us because those sites are, by our measure, poor quality sites for the vireo. The number of sites vireo occurs by itself in each habitat interested us because we assumed that these sites were high quality for the vireo. For each quantity we used the observed (i.e., na\"ive) occurrence state, rather than the latent state because we needed to compare each statistic to observed data. Comparing the latent states to observed data would have produced uninterpretable and large $P$--values. These six test statistics were predicted using each model, then compared to the observed values (i.e., the na\"ive co-occurrence state by habitat, and the na\"ive occurrence state of vireos without a cowbird by habitat) for each iteration. We used the comparisons to estimate the Bayesian $P$--value for each quantity and model. The Bayesian $P$--value is the probability that the predicted value was greater than or equal to the observed values \citep{gelmanEA2013}. $P$--values greater than 0.5 indicate that the model over predicts the quantity of interest, and low values suggest the model under predicts the quantity of interest. 

After this model fitting and selection processes, we estimated posterior predictive distributions for the probability of cowbird occurrence given vireo presence and the habitat specific vireo occurrence probability in a Bayesian framework with package R2OpenBUGS for program \textsf{R} \citep{openbugs,r}. R2OpenBUGS calls OpenBUGS from \textsf{R}. OpenBUGS is the open source iteration of the popular program WinBUGS, which uses a Gibbs sampling algorithm to generate samples from a posterior distribution. We used weakly informative priors for each parameter, meaning that each real parameter was given a Uniform(0,1) prior distribution. The only exception were the $\alpha 1$ parameters in equations \ref{eq:nai} and \ref{eq:z}, which were given a Normal(0,100) distribution. We ran each model with three chains, for 20,000 iterations each, thinning by 5. Because R2OpenBUGS discards the first half of each chain by default, the number of iterations kept for each chain was 2,000. We assessed convergence visually (e.g., with traceplots) and using the potential scale reduction factor \citep{gelmanrubin1992}. 

\section*{Results}

The probability of occurrence for vireos was higher in shade coffee, $\psi=0.71$ (95\% Bayesian credible interval: 0.51, 0.91) and forest, 0.66 (0.47, 0.87) than in sun coffee, 0.35 (0.19, 0.55;  Figure \ref{fig:co}). The probability of occurrence for cowbirds given vireo presence was higher in shade coffee, 0.92 (0.71, 0.99) and sun coffee, 0.90 (0.64, 0.99) than in forest, 0.28 (0.10, 0.55). The probability of co--occurrence between the cowbird and vireo was higher in shade coffee, 0.61 (95\% Bayesian credible interval: 0.40, 0.83), than in forest, 0.25 (0.14, 0.40), or sun coffee, 0.32 (0.16, 0.51). These estimates were derived from the best performing models of cowbird occurrence and vireo occurrence (Figures \ref{fig:modsel} and \ref{fig:ppc}; see below). The two species were predicted to co--occur at 7 of our sample sites in forest (95\% Bayesian credible interval: 4,12), 26 of our sample sites in shaded--coffee plantations (18, 33), and 12 sites in sun coffee (8, 16; Figure \ref{fig:ppd}). The vireo was predicted to occur without cowbirds at 21 of our sample sites in forests (12, 28), one site in shade coffee (0, 7), and one site in sun coffee (0, 3; Figure \ref{fig:ppd}).

Although no model of cowbird occurrence was the definitive favorite, the two--stage model was selected as the best performing model of cowbird occurrence by DIC and posterior predictive checks (Figures \ref{fig:modsel}, \ref{fig:ppc}, and \ref{fig:ppcHist}; Table \ref{tab:cowmods}). The two--stage model had the lowest DIC (844) of any cowbird occurrence model. It had the second highest $R^2$ value, 0.432, trailing only the best performing model that lacked biological interactions, 0.455. Models that included biological interactions better predicted the observed na\"ive co--occurrence state than the no-interaction model (Figure \ref{fig:ppc}). The two--stage model did the best job of predicting sites with a vireo detection and no cowbird detections. (Figures \ref{fig:ppc} and \ref{fig:ppcHist}). The base model for cowbird occurrence was the model that included the land use covariate affecting occurrence and the observer covariate affecting detection (Table \ref{tab:cowmods}) This model had the lowest DIC (378). The next best performing model, in terms of DIC (385), was the model with constant detection and land use induced heterogeneity in occurrence. 

For vireos, we selected the model with the land use covariate affecting occurrence and the observer covariate affecting detection (Table \ref{tab:virmods}). (This model was used to predict $z^V$ in the two--stage model and the conditional occurrence model.) While the $\psi(.) p(\text{observer})$ model had the lowest DIC (471), the model that included the land use and observer effects trailed only slightly (477). Further, the data suggest that vireo occurrence is affected by land use. The probability of occurrence in sun coffee plantations was considerably lower than in shade coffee plantations, and forests. There was little difference between vireo occurrence in shade-coffee plantations and forests. 

\section*{Discussion}

\begin{figure}[h!]
	\caption{Plot of posterior means, 50\%, and 95\% credible intervals for predicted probabilities of occurrence for a brood parasite, the Shiny Cowbird, and its host, the Puerto Rican Vireo, in each land use. The top plot shows the probability of occurrence for the vireo in each land use. The next plot shows the probability of cowbird occurrence given vireo occurrence. The last plot shows the 	probability of co--occurrence for the two species. The cowbird occurrence probabilities were estimated using the two--stage model. }
	\begin{center}
		\includegraphics[scale=1]{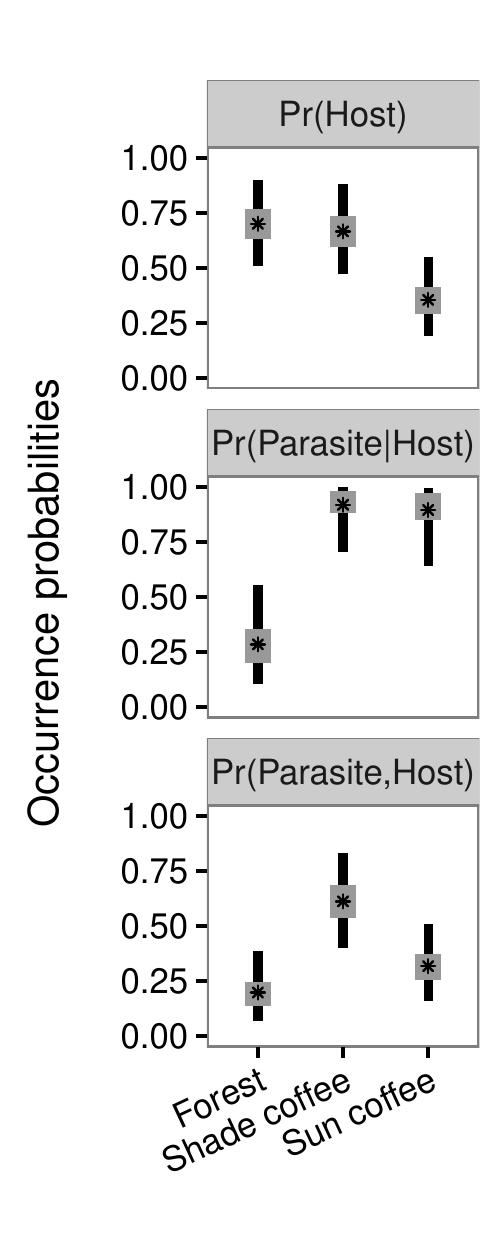} 
	\end{center}
	\label{fig:co}
\end{figure}

\begin{figure}[h!]
	\caption{Density plot of posterior predictive distributions of the predicted number of sites in each land use where cowbirds and vireos co-occur, and the predicted number of sites in each land use where the vireo occurs without cowbirds. In this case, we estimate the true occurrence state, rather than the observed occurrence state. Shade denotes shaded coffee plantations, and sun denotes sun coffee plantations. The distributions were generated using the two--stage model. }
	\label{fig:ppd}
	\begin{center}
		\includegraphics[width=71mm]{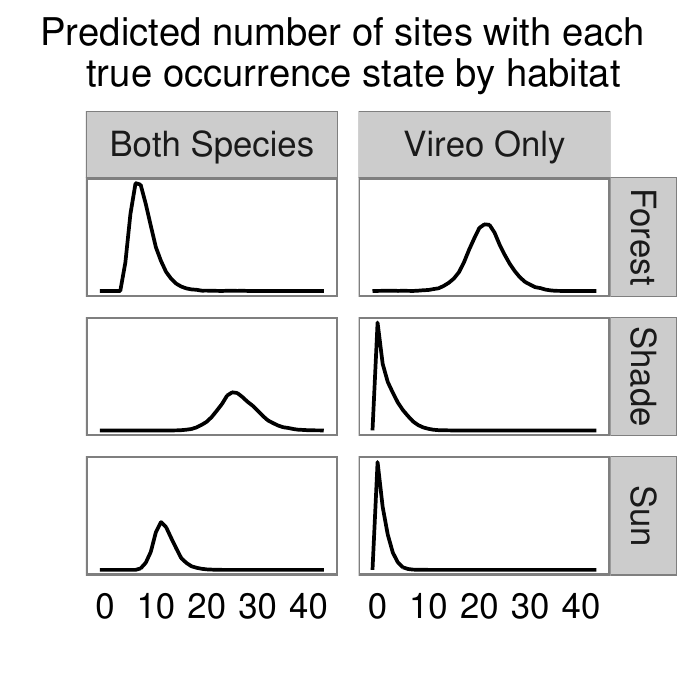}
	\end{center}
\end{figure}

We were interested in modeling habitat quality and habitat selection response for an endemic bird in an agroecological region of Puerto Rico, with the goal of providing information about the ecological value of particular land uses: forests, sun coffee plantations, and shaded--coffee plantations. We used occupancy as a measure of Puerto Rican Vireo landscape--level habitat selection, and the probability of Shiny Cowbird occurrence given vireo presence as a measure of habitat quality. Vireos seemed to select forests and shaded coffee plantations preferentially over sun coffee (Figure \ref{fig:co}). Both types of coffee plantations were poor quality habitat for the vireo, given the high probability of cowbird occurrence given vireo presence (Figure \ref{fig:co}). These two findings suggest that for vireos, forests are sources, sun coffee plantations are sinks, and shaded coffee plantations are ecological traps \citep[equal--preference traps in the parlance of][meaning that the trap is equally preferred to the source habitat]{robertsonEA2013}. While these results are not conclusive, we provided evidence that there are high rates of nest parasitism in coffee plantations, and that the vireo is preferentially occupying shade coffee plantations. 

In this paper, we used occupancy probability as a measure of habitat selection. We acknowledge that this in an imperfect measure. Ideally, estimates of habitat selection would utilize data on the size and location of home ranges (e.g., via telemetry) and how resources in the home range are utilized (e.g., via observations of marked individuals). Collecting these data for the vireo, a small songbird, would be incredibly resource intensive relative to our study. Our approach, in contrast, is relatively cheap and produces a sensible surrogate for habitat selection at the landscape scale. Occupancy probability can be interpreted as the probability that a home range overlaps the study site \citep{efforddawson2012}. As occupancy increases, the probability that a home range overlaps a site increases. Thus, if a species has a high occupancy probability at sites in one land use, then many sites in that that land use will have an overlapping home range, and there are many home ranges in that land use generally. This suggest that the species is selecting that land use at the landscape scale. Estimating habitat selection in this way is cheap and, when coupled with a measure of habitat quality, can effectively inform management. 

We used the probability of cowbird occurrence given vireo presence to infer vireo habitat quality from our data. There are better ways to estimate the quality of a particular habitat for a population or metapopulation. These include survival from mark--recapture data \citep{breiningerEA2009}, nest success from nest monitoring data \citep{tewksburyEA2006}, or overall population growth using both \citep{franklinEA2000}. Capturing, marking, and observing small passerines, or locating and revisiting nests, requires a significant amount of effort. Managers and wildlife biologists typically lack the resources to do such demanding and narrowly focused studies, unless doing so is in their mandate. Monitoring the presence/absence (honestly, detection/non--detection) of species in a community, conversely, is cheap and produces a versatile data set. We showed how this type of data, coupled with prior knowledge of the system's ecology, could be used to infer habitat quality. 

Future research could show that the probability of cowbird occurrence given vireo presence is not a good proxy of a more natural measure of habitat quality, such as vireo nest survival or population growth. Even if that is the case, this conditional probability provides relevant information. Using our example, high co--occurrence rates between the invasive cowbird and endemic vireo might discomfort managers, even if there isn't evidence of demographic consequences. 

We hypothesized that the presence of a vireo at a site may increase the probability that a cowbird occurs at a site. Cowbirds rely on quality hosts to breed; \cite{wiley1988} showed that vireo nests produce cowbird young as well or better than any other Puerto Rican songbird. Given that breeding events in birds are brief, one could think that this dependence in occurrence would be lost over the course of an occupancy study. However, cowbirds have been observed lingering around the parasitized nest until their young are fully fledged \citep{hooverrobinson2007}. This is one explanation for the improved performance of cowbird occurrence models that included information about the vireo occurrence state. We hypothesized that this relationship could arise in two different ways. In the first, which corresponded to the two--stage model, the presence of a vireo increased the probability of occurrence of a cowbird to the same degree regardless of land use. This model failed to account for the possibility that the presence of a vireo would affect cowbird occurrence less in land uses where cowbirds would occur anyway, such as sun coffee, an open land use. Despite lacking this nuance, which seemed to be supported by the raw data (Table \ref{tab:raw}), the two--stage model performed better than the conditional occurrence model (Figures \ref{fig:modsel} and \ref{fig:ppc}). The conditional model may have performed poorly because this model has many parameters. These types of models tend to better represent reality; however, they may perform worse than simpler and less realistic models when confronted with data \citep{hilbornmangel1997}. We also hypothesized that the two--stage model would perform better than the na\"ive model because the two stage model incorporates the estimated true occurrence state rather than the observed (i.e., na\"ive) occurrence state. The data supported this hypothesis. The two--stage model had a lower DIC, higher $R^2$, and performed better by posterior predictive checks. That said, these differences were slight. 

Our results suggest that, in our study area, shaded coffee plantations are an ecological trap for the Puerto Rican Vireo. In our study area, the proportion of shade coffee to sun coffee, or shade coffee to forest, is unknown; however, it is unlikely to be so high as to cause metapopulation decline for the vireo \citep{battin2004}. That said, it is unlikely that the vireo will quickly respond to a drop in habitat quality created by the cowbird because the vireo lacks evolutionary mechanisms for defending against exotic species \citep{battin2004}. Fortunately, our results and the results of \cite{tossas2008} suggest that the vireo still occurs with high probability in the good quality habitat in Puerto Rico's montane forests. If the coverage of this habitat remains stable, or expands, the vireo should have a large amount of habitat to safely reproduce in.

Shaded--coffee plantations may be an ecological trap for other songbird species in other study areas. We suspect that nest parasites or open habitat affiliated nest predators may be attracted to shaded--coffee plantations, as the Shiny Cowbird was in our example. There is a pressing need to understand whether this land use, which can provide ecological value for some species and economic value for local communities, subtly is an ecological trap. We hope that researchers will direct their attention to this issue. 

\section*{Acknowledgements}
This research was ultimately possible because of our colleagues, our technicians, private landowners, and our funders. We thank Katy Battle for being integral to the data collection process, and the Puerto Rico Department of Natural and Environmental Resources for allowing us to sample on lands they manage. Beatriz Romero, Luis Rodrigues, Damien Hardgrove, Joshua Morel, Mel Rivera, and Ray Robles were jovial, kind, and knowledgable technicians. Permission to sample from coffee growers was absolutely critical to our research. We were glad to receive their welcome each morning of the field season. This research was funded by the US Fish and Wildlife Service, the US Geological Survey, and the Puerto Rico Departmental of Natural and Environmental Resources. 

\bibliographystyle{jae}
\bibliography{ref}

\newpage
\section*{Figures}

\newpage

\newpage

\newpage

\begin{figure}[h!]
	\caption{Results from model selection of the three models that included biological interactions and the best performing model that assumed no interaction. Lower values of DIC suggest greater predictive ability; differences in DIC larger than 10 suggest significant differences among models. The $R^2$ presented is \cite{tjur2009}'s pseudo--$R^2$ for logistic regression. There were only subtle differences among models, with the no--interaction model performing the best.}
	\label{fig:modsel}
	\begin{center}
		\includegraphics[width=71mm]{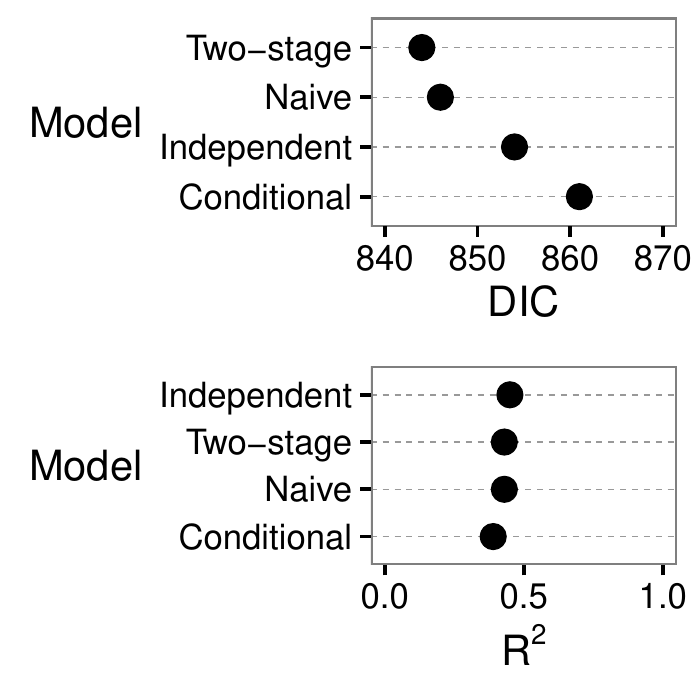}
	\end{center}
\end{figure}

\newpage

\begin{figure}[h!]
	\caption{Bayesian $P$--values from posterior predictive checks of the three models that included biological interactions and the best performing model that assumed no interaction. The test quantities of interest were the number of sites in each land use where the vireo and the cowbird are detected, and the number of sites in each land use where only the vireo is detected. The Bayesian $P$--value is the probability that the predicted value exceeds the observed values.}
	\label{fig:ppc}
	\begin{center}
		\includegraphics[width=71mm]{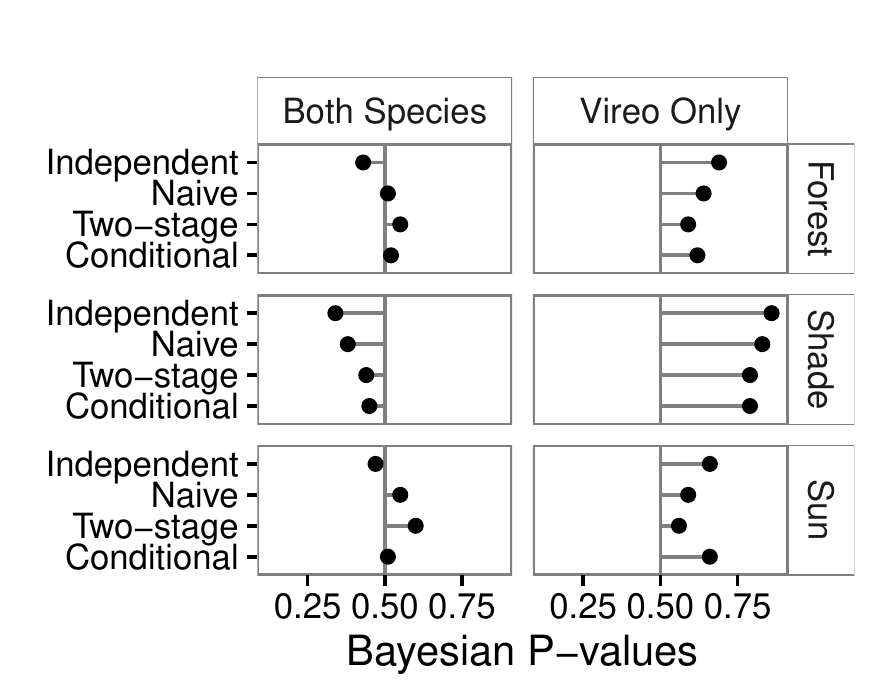}
	\end{center}
\end{figure}

\newpage

\section*{Tables}

\begin{table}[ht]
		\caption{Models fit to the cowbird detection/non--detection data. Independent models assume that the cowbird occurs independently of the vireo; dependent models assume that cowbird occurrence depends on vireo occurrence. The models in the righthand column are sometimes called (e.g., in Figures 2 and 3) the na\"ive model, the two--stage model, and the conditional model, respectively. The last model in the left hand--column is called the "no interaction" model in Figures 2 and 3.}
		\label{tab:mods}
		\begin{tabular}{l l l l}
		\hline
		\multicolumn{2}{c}{Independent Models} & \multicolumn{2}{c}{Dependent Models} \\  
		\hline
		Occurrence & Detection & Occurrence & Detection  \\  
		\hline
		 $\psi$(.) & $p$(.)& $\psi$(land use + $\mathrm{vireo}$) & $p$(observer)  \\ 
		 $\psi$(land use) & $p$(.) & $\psi$(land use + $\tilde{z}^H$) & $p$(observer) \\ 
		 $\psi$(.) & $p$(observer) & $\psi$\big(land use$*\tilde{z}^H$ + land use$*(1-\tilde{z}^H)\big)$ &  $p$(observer) \\ 
		 $\psi$(land use) & $p$(observer)  \\ 
		\hline
		\end{tabular}
\end{table}

\newpage

\begin{table}[h!]
	\caption{ Posterior means, 95\% credible intervals, DIC's, and coefficient of discrimination for the best performing independent model and each of the models including dependent interactions. ``Land'' refers to the land use covariate.}
	\centering
	\label{tab:biotic}
	\begin{tabular}{ l p{3.3cm} ccc}
		~\\ \hline
		Model & Parameter & Estimate & 2.5\% & 97.5\%  \\ \hline
		\multirow{5}{*}{\parbox{3cm}{Independent}}  
		& $p$ Obs. A    			 & 0.218 	& 0.144 		& 0.316  \\
		& $p$ Obs. B                 & 0.387    & 0.282         & 0.510      \\
		& $\psi$ in Forest           & 0.203    & 0.082         & 0.399    \\
		& $\psi$ in Sun              & 0.774    & 0.526         & 0.978   \\
		& $\psi$ in Shade            & 0.849    & 0.623         & 0.991   \\ \hline
		\multirow{6}{*}{\parbox{2.8cm}{Na\"ive}}         
		& $p$ Obs. A                 & 0.215    & 0.142         & 0.311  \\
		& $p$ Obs. B                 & 0.380    & 0.278         & 0.502      \\
		& $\psi$ in Forest           & 0.102    & 0.016         & 0.314   \\
		& $\psi$ in Sun              & 0.719    & 0.455         & 0.965    \\
		& $\psi$ in Shade            & 0.816    & 0.522         & 0.989  \\
		& $\alpha1_{y^H_{j\bullet}}$ & 1.292    & -0.232        & 3.471     \\ \hline
		\multirow{6}{*}{\parbox{2.8cm}{Two--stage}} 
		& $p$ Obs. A                 & 0.214    & 0.141         & 0.310  \\
		& $p$ Obs. B                 & 0.380    & 0.276         & 0.502   \\
		& $\psi$ in Forest           & 0.073    & 0.004         & 0.313   \\
		& $\psi$ in Sun              & 0.704    & 0.407         & 0.962     \\
		& $\psi$ in Shade            & 0.782    & 0.393         & 0.987      \\
		& $\alpha1_{\hat{z}^H_j}$          & 1.559    & -0.350        & 4.751   \\ \hline
		\multirow{8}{*}{\parbox{3cm}{Conditional}}
		& $p$ Obs. A                 & 0.228    & 0.150         & 0.326  \\
		& $p$ Obs. B                 & 0.402    & 0.294         & 0.525          \\
		& $\psi$ (Forest with)       & 0.243    & 0.090         & 0.490        \\
		& $\psi$ (Forest without)    & 0.141    & 0.007         & 0.603      \\
		& $\psi$ (Sun with)          & 0.859    & 0.521         & 0.994      \\
		& $\psi$ (Sun without)       & 0.640    & 0.336         & 0.943      \\
		& $\psi$ (Shade with)        & 0.836    & 0.559         & 0.991      \\
		& $\psi$ (Shade without)     & 0.755    & 0.272         & 0.986        \\
	\end{tabular}
\end{table}

\newpage

\begin{table}[h!]
	\caption{Posterior means, 95\% credible intervals, and DIC's for each of the four cowbird models that do not include dependent interactions. }
	\label{tab:cowmods}
	\begin{center}
		\begin{tabular}{l  l cccc}
			~\\
			
			Model                                          & Parameter        & Estimate & Lower (2.5\%) & Upper (97.5\%) & DIC                    \\ \hline
			\multirow{2}{*}{$\psi$(.) p(.)}                & $p$              & 0.280    & 0.189         & 0.385          & \multirow{2}{*}{546.4}  \\ 
			& $\psi$            & 0.641    & 0.467         & 0.900          &                        \\ \hline
			\multirow{3}{*}{$\psi$(.) $p$(observer)}       & $p$ (Obs. A)       & 0.212    & 0.129         & 0.322          & \multirow{3}{*}{514.4} \\ 
			& $p$ (Obs. B)       & 0.374    & 0.250         & 0.514          &                        \\
			& $\psi$           & 0.623    & 0.460         & 0.864          &                        \\ \hline
			\multirow{4}{*}{$\psi$(land use) $p$(.)}        & $p$              & 0.294    & 0.222         & 0.387          & \multirow{4}{*}{384.8} \\
			& $\psi$ (Forest) & 0.206    & 0.084         & 0.407          &                        \\
			& $\psi$ (Sun)    & 0.780    & 0.533         & 0.980          &                        \\
			& $\psi$ (Shade)  & 0.851    & 0.620         & 0.991          &                        \\ \hline
			\multirow{5}{*}{\parbox{3cm}{$\psi$(land use) $p$(observer)}} & $p$ (Obs. A)       & 0.218    & 0.144         & 0.316          & \multirow{5}{*}{377.6} \\
			& $p$ (Obs. B)     & 0.387    & 0.282         & 0.510          &                        \\
			& $\psi$ (Forest) & 0.203    & 0.082         & 0.399          &                        \\
			& $\psi$ (Sun)    & 0.774    & 0.526         & 0.978          &                        \\
			& $\psi$ (Shade)  & 0.849    & 0.623         & 0.991          &           \\ \hline    
		\end{tabular}
	\end{center}
\end{table}

\newpage

\begin{table}[h!]
	\caption{Posterior means, credible intervals, and DIC's for the four Vireo models.}
	\centering
	\label{tab:virmods}
	\begin{tabular}{l  l cccc}
		~ \\
		Model                                          & Parameter        & Estimate & Lower (2.5\%) & Upper (97.5\%) & DIC                    \\ \hline
		\multirow{2}{*}{$\psi$(.) p(.)}                & $p$              & 0.414    & 0.320         & 0.509          & \multirow{2}{*}{478.5} \\
		& $\psi$            & 0.577    & 0.457         & 0.729          &                        \\ \hline
		\multirow{3}{*}{$\psi$(.) $p$(observer)}       & $p$ (Obs. A)       & 0.386    & 0.281         & 0.500          & \multirow{3}{*}{471.2} \\
		& $p$ (Obs. B)       & 0.454    & 0.332         & 0.577          &                        \\
		& $\psi$           & 0.573    & 0.452         & 0.718          &                        \\ \hline
		\multirow{4}{*}{$\psi$(land use) $p$(.)}        & $p$              & 0.407    & 0.315         & 0.505          & \multirow{4}{*}{491.4} \\
		& $\psi$ (Forest) & 0.717    & 0.512         & 0.926          &                        \\
		& $\psi$ (Sun)    & 0.353    & 0.194         & 0.562          &                        \\
		& $\psi$ (Shade)  & 0.666    & 0.470         & 0.881          &                        \\ \hline
		\multirow{5}{*}{\parbox{3cm}{$\psi$(land use) $p$(observer)}} & $p$ (Obs. A)    & 0.379    & 0.276         & 0.492          & \multirow{5}{*}{477.1} \\
		& $p$ (Obs. B)       & 0.449    & 0.331         & 0.576          &                        \\
		& $\psi$ (Forest) & 0.709    & 0.513         & 0.915          &                        \\
		& $\psi$ (Sun)   & 0.348    & 0.192         & 0.551          &                        \\
		& $\psi$ (Shade)  & 0.657    & 0.468         &   0.869             &                       
	\end{tabular}
\end{table}

\newpage

\begin{table}[ht]
	\centering
	\caption{Number of sites with and without a vireo detection, separated by land use and na\"ive cowbird occurrence state.}
	\label{tab:raw}
	\begin{tabular}{| l | rr | rr | }
		\hline 
		& \multicolumn{2}{c|}{Vireo Detection} & \multicolumn{2}{c|}{No Vireo Detection} \\ \hline
		Habitat & Cowbird   & No Cowbird   & Cowbird     & No Cowbird    \\ \hline
		Forest  & 4         & 19           & 1           & 17            \\
		Sun     & 7         & 3            & 12          & 15            \\
		Shade   & 13        & 9            & 11          & 9      \\      \hline 
	\end{tabular}
\end{table}

\end{document}